\documentclass[prb, onecolumn, reprint, superscriptaddress, aps, longbibliography, floatfix]{revtex4-2}

\usepackage[utf8]{inputenc}
\usepackage[T1]{fontenc}

\usepackage{graphicx}
\usepackage{amsmath}
\usepackage{amssymb}
\usepackage{afterpage}
\usepackage{amsfonts}
\usepackage{braket}
\usepackage{float}
\usepackage{bm}
\usepackage{siunitx}
\usepackage{lipsum}
\usepackage[]{xcolor}
\usepackage[english]{babel}
\usepackage{hyperref}
\usepackage{physics}
\hypersetup{
	colorlinks,
	linkcolor={blue!90!black},
	citecolor={blue!90!black},
	urlcolor=	{blue!90!black}
}
\usepackage[]{placeins}
\usepackage{makecell}

\usepackage{orcidlink}

\usepackage[]{newtxtext} %arxiv OFF
\usepackage[subscriptcorrection,nosymbolsc,smallerops,bigdelims]{newtxmath} %arxiv OFF
\DeclareMathAlphabet{\mathcal}{OMS}{cmsy}{m}{n} %arxiv OFF
\DeclareMathAlphabet{\mathbcal}{OMS}{cmsy}{b}{n} %arxiv OFF
\begin{document}

\title{Phonon polariton confinement in isotopically pure MOVPE-grown BN triangles}

\author{Maximilian~Scharpey\,\orcidlink{0009-0006-3642-6830}}
\affiliation{Institute of Physics, University of M\"unster, M\"unster, Germany}

\author{Oskar~Schr\"oer\,\orcidlink{0009-0006-5707-6569}}
\affiliation{Institute of Physics, University of M\"unster, M\"unster, Germany}

\author{Daniel Wigger\,\orcidlink{0000-0002-4190-8803}}
\affiliation{Department of Physics, University of M\"unster, Germany}

\author{Lena~Miler\,\orcidlink{0009-0002-3676-8129}}
\affiliation{Faculty of Physics, University of Warsaw, Warsaw, Poland}

\author{Jakub~Iwa{\'n}ski\,\orcidlink{0000-0003-2395-4010}}
\affiliation{Faculty of Physics, University of Warsaw, Warsaw, Poland}

\author{Andrzej~Wysmo\l{}ek\,\orcidlink{0000-0002-8302-2189}}
\affiliation{Faculty of Physics, University of Warsaw, Warsaw, Poland}

\author{Johannes~Binder\,\orcidlink{0000-0002-0461-7716}}
\affiliation{Faculty of Physics, University of Warsaw, Warsaw, Poland}

\author{Iris~Niehues\,\orcidlink{0000-0001-7438-2679}}
\email{iris.niehues@uni-muenster.de}
\affiliation{Institute of Physics, University of M\"unster, M\"unster, Germany}

%\email{email} 

\begin{abstract}
Phonon polaritons, quasiparticles formed by the resonant hybridization of light and lattice vibrations, exhibit unique properties like the possibility of hyperbolic dispersions. In this context, exfoliated hexagonal boron nitride (hBN) has emerged as a promising material for phonon polariton-based research. However, to advance toward practical applications, it is essential to demonstrate efficient phonon-polariton propagation and confinement in large-area epitaxial BN. To address this topic, we use metalorganic vapor phase epitaxy (MOVPE)-grown BN and investigate the phonon polariton properties using scattering-type scanning near-field optical microscopy (s-SNOM) and nanoscale Fourier transform infrared spectroscopy (nano-FTIR). We report remarkably long phonon polariton propagation lengths, indicating the high crystalline quality of the BN layer.  By using epitaxially grown isotopically pure triangular islands, we further demonstrate efficient phonon-polariton confinement with mode patterns tunable by the incident light wavelength. Our results pave the way for implementing all-epitaxial, microscale, high-quality polariton resonators for nanophotonics and quantum optics.
\end{abstract}

\date{\today}

\maketitle

\section{Introduction}\label{sec:intro}
In recent years, hexagonal boron nitride (hBN) has emerged as a highly promising material for nanophotonic applications, among other reasons, due to its ability to support phonon polaritons with a hyperbolic dispersion~\cite{caldwell2014sub,yoxall2015direct,wang2024planar}. These quasiparticles, which arise from the resonant hybridization of electromagnetic waves (photons) and lattice vibrations (phonons) in the material, exhibit unique properties such as strong confinement and minimal propagation losses. Using scattering-type scanning near-field optical microscopy (s-SNOM), one can successfully launch and image phonon polaritons in hBN with nanoscale resolution~\cite{dai2014tunable,hillenbrand2025visible}.\\
While previous studies have primarily focused on polaritons in exfoliated hexagonal boron nitride (hBN), relatively little research has been conducted on hBN, which was grown using chemical vapor deposition (CVD) or metalorganic vapor phase epitaxy (MOVPE). A significant advantage of CVD and MOVPE growth techniques is their ability to produce large-area samples of hBN\cite{chugh2018flow,moon2023hexagonal,tatarczak2026deterministic,li2016large,tokarczyk2023effective,wang2024bevel} relevant for technological applications. Moreover, these methods allow to taylor the crystal properties to a certain degree by choosing the growth parameters accordingly. On the one hand synthesis of different lattice structures is possible~\cite{wang2024bevel,iwanski2024revealing}. On the other hand the use of isotopically pure materials can be targeted~\cite{janzen2024boron,liu2018single,vuong2018isotope}.
Here, we focus on investigating epitaxial natural boron nitride ($^{\text{nat}}$BN) and triangular isotopically pure BN islands of  $^{10}$BN grown using MOVPE.  To judge the overall quality of the growth process, the density of defects is a key metric, which directly impacts the propagation properties of polaritons in these samples, as highlighted in recent studies on isotopically pure~\cite{giles2018ultralow} and on CVD grown natural BN~\cite{calandrini2023near}. Our findings reveal remarkably long phonon polariton propagation lengths, indicating the high quality of the material. Notably, we demonstrate that by growing triangular islands of  $^{10}$BN on $^{\text{nat}}$BN layers we can achieve polariton confinement in microscaled geometries. Until now such confinement was achieved by hBN structuring or structuring of the substrate~\cite{alfaro2017nanoimaging,autore2018boron,dai2018hyperbolic,dai2018manipulation,chaudhary2019polariton,ramer2020high,li2020collective,duan2022active,herzig2024high,jackering2025tailoring,borodin2026probing,borodin2026cavity} or by hand-made stacking of different layers to produce heterostructures~\cite{chen2023van}. We show here that we can control the polariton propagation in our samples directly by tayloring the growth process. By modeling the confined polariton modes, we are able to reproduce the spatial field distributions in amplitude and phase. This allows to extract the dispersion relation of the polaritons in the triangular islands of $^{10}$BN.

\section{Results and discussion}\label{sec:results}
\subsection{Polariton propagation properties}
%%%
\begin{figure}[t]
\centering
\includegraphics[width=\columnwidth]{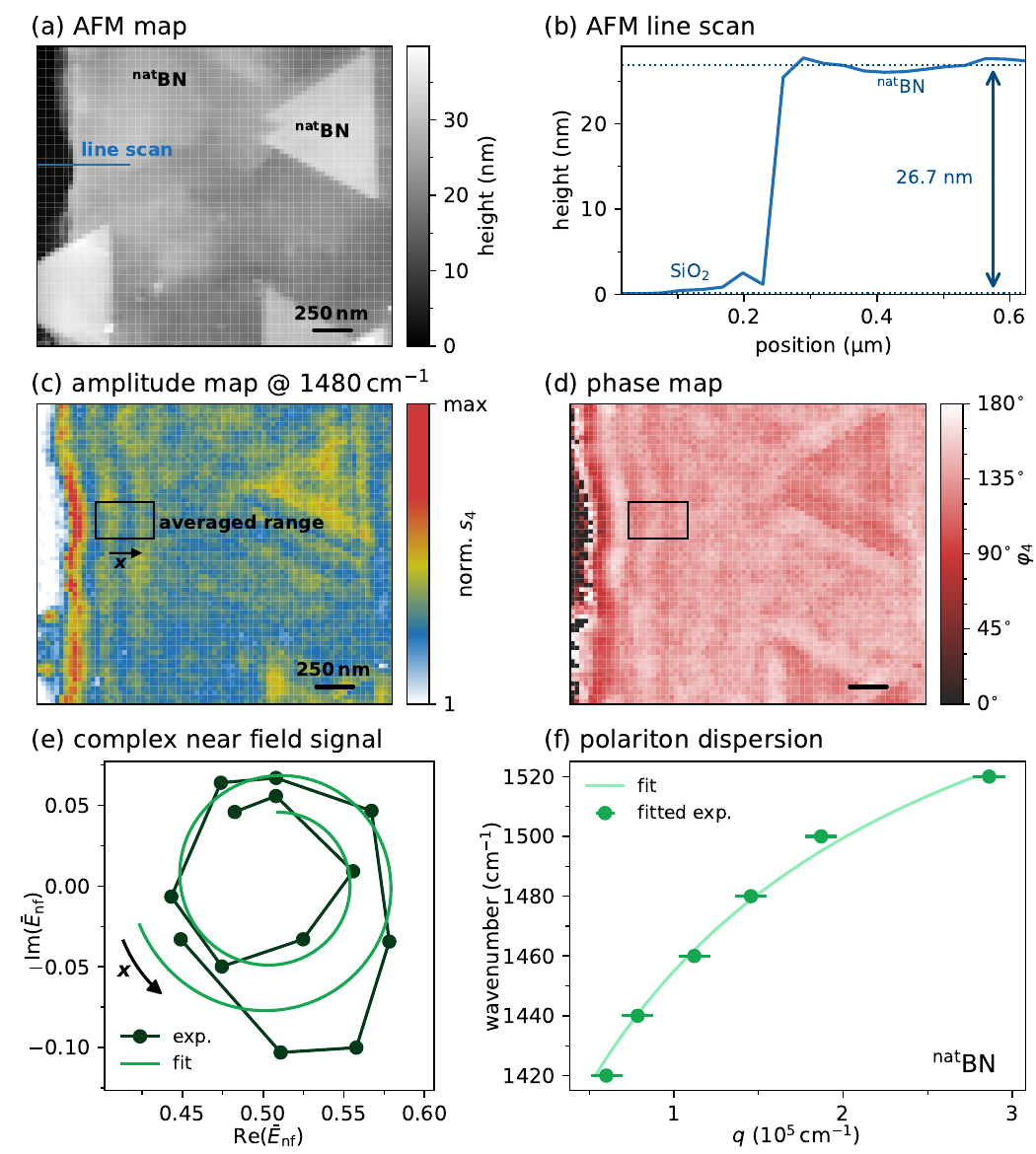}
\caption{Phonon polariton characterization of MOVPE grown $^{\text{nat}}$BN. (a) AFM topography map of the sample also showing $^{\text{nat}}$BN triangles (see Figure~\ref{fig:2}). (b) AFM line scan across the edge of the $^{\text{nat}}$BN layer. (c) s-SNOM amplitude $s_4$ (fourth demodulation order) of the sample area from (a) for an excitation wavenumber of 1480~cm$^{-1}$. (d) s-SNOM phase map $\varphi_4$ corresponding to (c). (e) Complex valued s-SNOM signal $\bar{E}_{\rm nf}$ (dark green dots) as function of polariton propagation distance $x$ vertically averaged over the rectangle in (c, d). Fit (green line) according to Eq.~\eqref{eq:spiral} with resulting $q=1.45\times 10^5$~cm$^{-1}$. (f) Measured phonon polariton dispersion (green dots) with theoretical fit (light green line) according to Eq.~\eqref{eq:dispersion}. The resulting phonon wavenumbers are $\omega_{{\rm TO},x} = (1354\pm 8)$~cm$^{-1}$ and $\omega_{{\rm LO},x} = (1608\pm6)$~cm$^{-1}$.}\label{fig:1}
\end{figure}
%%%

We first investigate the polariton propagation in the MOVPE grown $^{\text{nat}}$BN layer using scattering-type scanning near-field optical microscopy (s-SNOM). For the s-SNOM measurements, we use a commercially available near-field microscope (Neascope) from Neaspec (attocube systems GmbH) equipped with a quantum cascade laser (1300~cm$^{-1}$-1700~cm$^{-1}$) covering the entire upper reststrahlen band of BN ($\approx 1370$~cm$^{-1}$-1610~cm$^{-1}$)~\cite{caldwell2014sub}. This near-field technique relies on the elastic light scattering at a metallic tip of an atomic force microscope (AFM), where the tip is illuminated with a focused laser beam and acts as an optical antenna, creating a highly localized near-field (nanofocus) at its apex~\cite{keilmann2004near, chen2019modern}. By interferometric detection of the backscattered light and demodulation with respect to the AFM's tapping frequency, background free optical amplitude and phase images with a spatial resolution of down to 20~nm can be recorded alongside the topography~\cite{ocelic2006pseudoheterodyne}. Note that in contrast to standard Neascopes, our setup includes a high-quality, silver-protected, off-axis parabolic mirror with a numerical aperture (NA) of 0.72, which optimizes the focusing and collection efficiency of the optical system~\cite{niehues2025nanoscale}. Furthermore, s-SNOM has been expanded to enable nanoscale infrared spectroscopy (nano-FTIR), i.e., hyperspectral nanoimaging, providing valuable insights into IR-active phonon modes and strain distributions at the nanoscale~\cite{huth2012nano} (see Supporting Information (SI) for more measurement details).\\
In previous studies, it was found that, in CVD grown hBN, one does not observe polariton fringes as in exfoliated hBN but a noisy pattern due to the high density of lattice defects, which lead to scattering and interference of the polaritons~\cite{calandrini2023near}. In contrast, in our MOVPE grown $^{\text{nat}}$BN samples shown in Figure~\ref{fig:1}(a) as AFM image, we observe clean polariton fringes (Figure~\ref{fig:1}(c,d)) only with a higher damping rate, i.e., a shorter propagation length, compared to their counterparts in exfoliated hBN~\cite{dai2014tunable}. Figure~\ref{fig:1} shows the polariton propagation characterization for this sample which consists of a 26.7~nm thick layer of $^{\text{nat}}$BN grown on sapphire and transferred onto a Si/SiO$_2$ substrate as described in Refs.~\cite{iwanski2021delamination,ludwiczak2024large}. The topography and a corresponding line scan are displayed in Figures~\ref{fig:1}(a,b). Figures~\ref{fig:1}(c,d) show the s-SNOM amplitude $s_4$ (fourth demodulation order) and phase $\varphi_4$ image taken at 1480~cm$^{-1}$, clearly showing the polariton fringes launched at the tip and reflected from the edge of the $^{\text{nat}}$BN layer. In addition to the flat $^{\text{nat}}$BN region the topography shows $^{\text{nat}}$BN triangular islands grown on top of the $^{\text{nat}}$BN continuous layer (see SI for more information). We will focus on similar but isotopically pure $^{10}$BN islands in Figures~\ref{fig:2}-\ref{fig:5}. To analyze the phonon polariton properties of the $^{\text{nat}}$BN layer, we select rectangular regions near the flake edge (avoiding edge artifacts~\cite{mester2022high}) where the wave fronts appear approximately straight. To reduce the noise of the signal we average over the height of the rectangle. The $x$-coordinate will be the propagation distance in the following. Combining s-SNOM amplitude and phase into one complex near field signal $\bar{E}_{\rm nf}$ we plot~\cite{yoxall2015direct}
\begin{align}
	\bar{E}_{\rm nf} &= s_4 \exp(i\varphi_4) = {\rm Re}(\bar{E}_{\rm nf}) + i {\rm Im}(\bar{E}_{\rm nf}) \label{eq:complex}
\end{align}
in Figure~\ref{fig:1}(e) as dark green points. We clearly see that the complex signal spirals inwards resembling the oscillating, decaying polariton wave. In order to retrieve the corresponding wave vector $q$, we fit (green line) the experimental data with the function
\begin{align}
	E_{\rm nf}^\text{(fit)} &= E_0 \exp[(i q -\gamma )2x] + E_\infty\label{eq:spiral}\ .
\end{align}
We perform this procedure for several excitation wavenumbers in the reststrahlen band between 1420 cm$^{-1}$ and 1520 cm$^{-1}$ (see SI for all measurements and fits) and plot the fitted data in Figure~\ref{fig:1}(f). We already see that the data follows the curved dispersion relation of phonon polaritons in flat hBN crystals confirming the high quality of our samples. The expected dispersion is given by~\cite{eremets1995optical,dai2014tunable}
\begin{subequations}\label{eq:dispersion}\begin{align}
	q(\omega) &= {\rm Re}\Bigg\{ i\frac{1}{d}\sqrt{\frac{\varepsilon_{z}(\omega)}{\varepsilon_{x}(\omega)}} \Bigg[  \arctan\left(\frac{i\varepsilon_{\text{air}}}{\sqrt{\varepsilon_{z}(\omega)\varepsilon_x(\omega)}}\right) \notag\\
		&\qquad\qquad + \arctan\left(\frac{i\varepsilon_{\text{SiO}_2}}{\sqrt{\varepsilon_{z}(\omega)\varepsilon_x(\omega)}}\right) \Bigg] \Bigg\} \\
	\varepsilon_n(\omega) &= \varepsilon_{n,\infty} \left[1+\frac{\omega_{{\rm LO},n}^2-\omega_{{\rm TO},n}^2}{\omega_{{\rm TO},n}^2-\omega^2-i\omega\gamma_n}\right]\,,\quad n=x,z
\end{align}\end{subequations}
We use established values for $^{\text{nat}}$BN for the out-off-plane (ZO-modes) phonon wavenumbers\cite{franke1997phase} $\omega_{{\rm TO},z} = 775$~cm$^{-1}$ and $\omega_{{\rm TO},z} = 820$~cm$^{-1}$, for the dielectric constants of the environment $\varepsilon_{\text{air}}=1$ and $\varepsilon_{\text{SiO}_2}=1.2$~\cite{kitamura2007optical}, high frequency dielectric constants~\cite{cai2007infrared} of BN $\varepsilon_{\infty,x}=4.87$ and $\varepsilon_{\infty,z}=2.95$, and the spectral broadenings~\cite{dai2014tunable} $\gamma_x=5$~cm$^{-1}$ and $\gamma_z=4$~cm$^{-1}$. For the layer thickness we use the value from the AFM scan in Figure~\ref{fig:1}(b), i.e., $d=26.7$~nm. Finally, we use the in-plane phonon wavenumbers as fitting parameters and retrieve the best agreement with the experiment for $\omega_{{\rm TO},x} = (1354\pm 8)$~cm$^{-1}$ and $\omega_{{\rm LO},x} = (1608\pm6)$~cm$^{-1}$. Typical literature values for these wavenumbers lie in the ranges~\cite{franke1997phase,ohba2001first,iwanski2023temperature,geick1966normal,segura2020long} $\omega_{{\rm TO},x}^{\rm (lit)} = 1367-1400$~cm$^{-1}$ and $\omega_{{\rm LO},x}^{\rm (lit)} = 1585-1610$~cm$^{-1}$. Our results perfectly agree with these values, which again confirms the high quality of our MOVPE-grown $^{\text{nat}}$BN crystals. Note, that variations in the literature phonon energies can potentially stem from strain~\cite{jha2014strain,androulidakis2018strained} due to the growth and transfer process and the quality of the material.

%%%%%%%
\subsection{Polariton confinement}
%%%
\begin{figure}[t]
\centering
\includegraphics[width=\columnwidth]{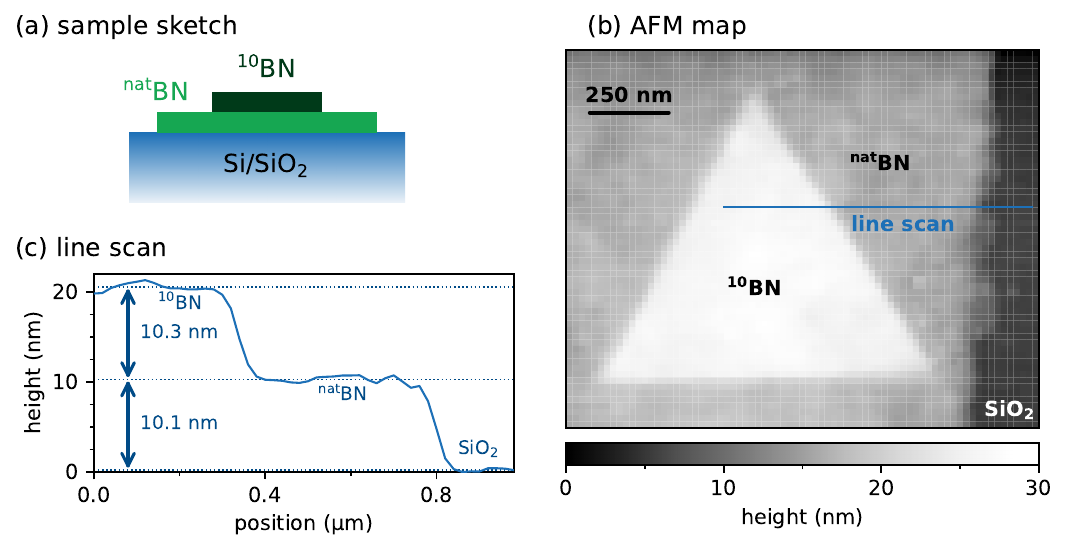}
\caption{Isotopically pure $^{10}$BN triangular islands. (a) Schematic of the sample structure consisting of a MOVPE-grown $^{\text{nat}}$BN layer with isotopically pure triangular islands ($^{10}$BN) transferred on a Si/SiO$_2$ substrate. (b) AFM topography image and (c) extracted line scan of the investigated area. The isotopically pure $^{10}$BN triangles with a thickness of about 10~nm are grown on a 10~nm thick $^{\text{nat}}$BN layer.}\label{fig:2}
\end{figure}
%%%
We now move our focus to the isotopically pure $^{10}$BN triangular islands, grown by MOVPE on the $^{\text{nat}}$BN layers. Our sample consists of 10~nm thick isotopically pure $^{10}$BN layers with triangular shapes on top of a 10~nm thick continuous layer of $^{\text{nat}}$BN~\cite{chugh2018flow,dabrowska2020two,tokarczyk2023effective,binder2023epitaxial} (see SI for more details). Note that we use a different sample here than in Figure~\ref{fig:1} as phonon polariton propagation studies in 10~nm thick $^{\text{nat}}$BN layers would be significantly more challenging due to the flat dispersion~\cite{dai2014tunable}. We use isotopically pure $^{10}$BN since it is known for hBN that the phonon polariton propagation length increases in such samples~\cite{giles2018ultralow}.
The triangular shape of the islands suggests that BN may adopt in a different sp$^2$-bonded polytype; however, here we focus on isotopic effects, which have a much stronger influence on the phonon properties~\cite{gilbert2019alternative,binder2023epitaxial,gil2022polytypes}.
The sample geometry of $^{10}$BN/$^{\text{nat}}$BN is shown in Figure~\ref{fig:2}(a), and an AFM topography image in (b). The blue line indicates the extracted line scan shown in (c) revealing the respective thicknesses of the investigated layers.

%%%
\begin{figure*}
\centering
\includegraphics[width=0.65\textwidth]{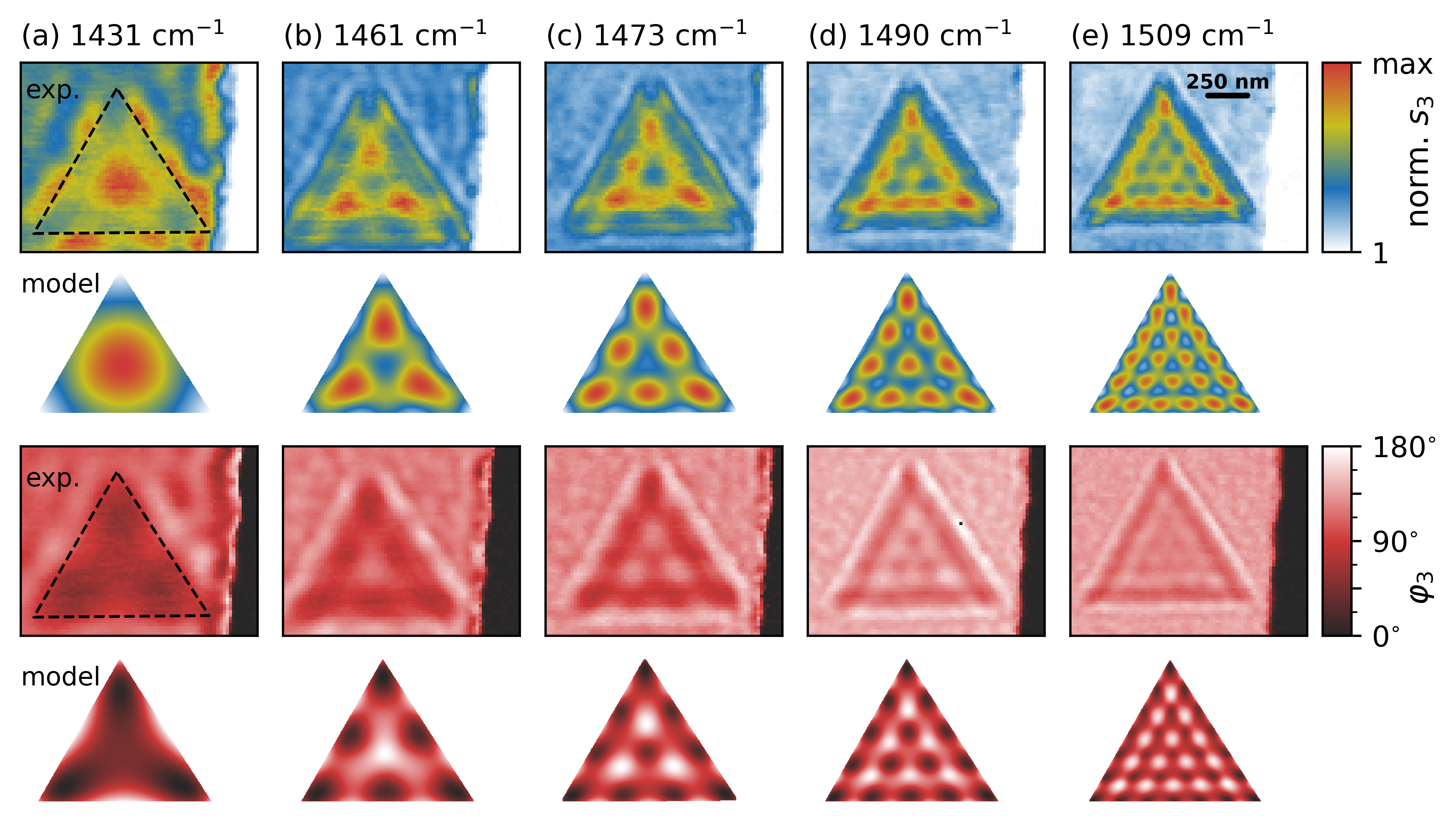}
\caption{Confined polariton modes in $^{10}$BN triangular islands. Upper two rows: s-SNOM $s_3$ amplitude images (3rd demodulation order) taken at selected excitation wavenumbers increasing from (a) to (e) and corresponding simulations. High intensities are shown in red and low intensities in blue colors. All images are initially normalized to the respective Si/SiO$_2$ substrate value and finally to their respective maximum value. Lower two rows: corresponding s-SNOM phase $\varphi_3$ images again normalized to the Si/SiO$_2$ substrate and simulations. The outline of the $^{10}$BN triangle is marked in (a) by the dashed black line.}\label{fig:3}
\end{figure*}
%%%

We again take s-SNOM images of these $^{10}$BN/$^{\text{nat}}$BN structures with different excitation wavenumbers in the upper Reststrahlen band. The results for amplitudes $s_3$ (top) and phases $\varphi_3$ (bottom) are shown in Figure~\ref{fig:3} for selected wavenumbers increasing from 1431~cm$^{-1}$ in (a) to 1509~cm$^{-1}$ in (e). In total we have recorded 123 s-SNOM images which can be found as a video in the SI. Looking first at the lowest wavenumber (energy) in (a), we find two distinct features: (I) The edges of the $^{10}$BN triangular island (marked as dashed black line) function as scattering edges for phonon polaritons propagating away from the triangle inside the $^{\text{nat}}$BN film below the $^{10}$BN island. We have already characterized the phonon polaritons in the epitaxial $^{\text{nat}}$BN in Figure~\ref{fig:1}. (II) Confined phonon polariton waves inside the $^{10}$BN triangle, which manifest as single maximum in the s-SNOM amplitude $s_3$. The rest of this article focuses on these phonon polaritons confined to the $^{10}$BN triangles. Note that until now polariton confinement was only achieved by hBN or substrate structuring~\cite{alfaro2017nanoimaging,autore2018boron,dai2018hyperbolic,dai2018manipulation,chaudhary2019polariton,ramer2020high,li2020collective,duan2022active,herzig2024high,jackering2025tailoring,borodin2026probing,borodin2026cavity}. Considering the phase image for the lowest wavenumber in (a, bottom) we find three distinct minima pointing towards the corners of the $^{10}$BN flake. In other words, we find two minima when scanning close to an edge of the triangle. A very similar pattern appears in the s-SNOM amplitude for the next higher depicted wavenumber in Figure~\ref{fig:3}(b, top). We find two maxima along any edge of the triangle. We have chosen this excitation wavenumber because it shows this similar distinct pattern as for $\varphi_3$ in Figure~\ref{fig:3}(a, bottom). Moving again from $s_3$ in (b, top) to the respective phase $\varphi_3$ in the bottom, the number of minima along each edge of the triangle increases from two to three. We continue this sequence of similar patterns of minima in $\varphi_3$ and maxima in $s_3$ for the next higher wavenumber followed by an increase in the number of extrema when moving from $s_3$ to $\varphi_3$ at the same wavenumber through Figures~\ref{fig:3}(c, d, e). The largest number of clear maxima along the edge in the amplitude (five) is found for 1509~cm$^{-1}$ in Figure~\ref{fig:3}(e). To model the clearly resolved maxima of the confined modes we have developed an interference model (described in the SI) that takes into account the phonon polariton creation via the tip. Those waves are reflected from the three edges of the triangle (with a phase shift of $\pi$) and then interfere with the local field of the tip's near field while being scattered out into the detected field. By choosing the wave vector $q$ to match the measured s-SNOM images, we achieve an excellent agreement between experiment and model for each depicted excitation frequency. A series of increasing frequency alongside the 123 measurements can be found as video in the SI as well as the extracted phonon dispersion relation for the $^{10}$BN crystal.

%%%
\begin{figure}
\centering
\includegraphics[width=0.45\textwidth]{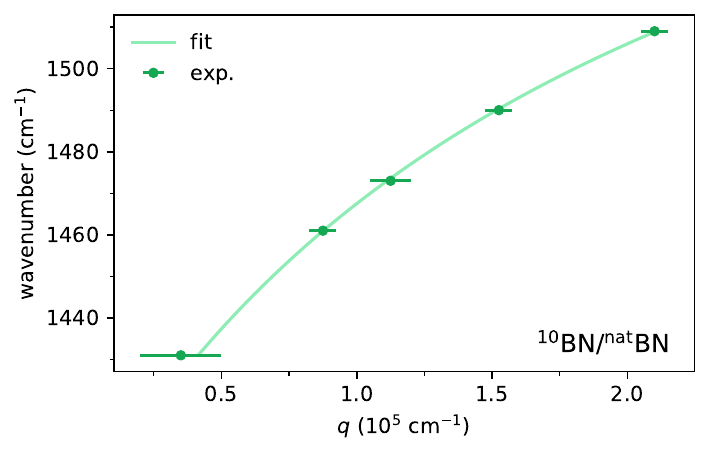}
\caption{Phonon polariton dispersion of the $^{10}$BN triangle determined from the confined modes in Figure~\ref{fig:3}. The resulting phonon wavenumbers are $\omega_{{\rm TO},x} = (1393\pm 6)$~cm$^{-1}$ and $\omega_{{\rm LO},x} = (1621\pm 7)$~cm$^{-1}$.}\label{fig:4}
\end{figure}
%%%

In Figure~\ref{fig:3} we have determined simulated triangle mode patterns that best reproduce the measured ones by variation of the polariton wave vectors $q$ at a fixed damping rate of $\gamma= 1.25$~\textmu m$^{-1}$. In Figure~\ref{fig:4} we plot these wave vectors together with the corresponding laser excitation wavenumbers. We can again fit this dispersion with the expected model from Eq.~\eqref{eq:dispersion}. We consider a layer thickness of $d=20$~nm, which includes the $^{10}$BN triangle itself and the $^{\text{nat}}$BN below. The best fit is found for the phonon wave numbers $\omega_{{\rm TO},x} = (1393\pm 6)$~cm$^{-1}$ and $\omega_{{\rm LO},x} = (1621\pm 7)$~cm$^{-1}$. This is again in excellent agreement with the expected phonon energies for isotopically pure $^{10}$BN (see Fig.~S1 in SI), where an overall blueshift around 30~cm$^{-1}$ has been reported~\cite{giles2018ultralow}.

%%%%%%%
\subsection{Spatio-spectral analysis}
%%%
\begin{figure*}
\centering
\includegraphics[width=0.6\textwidth]{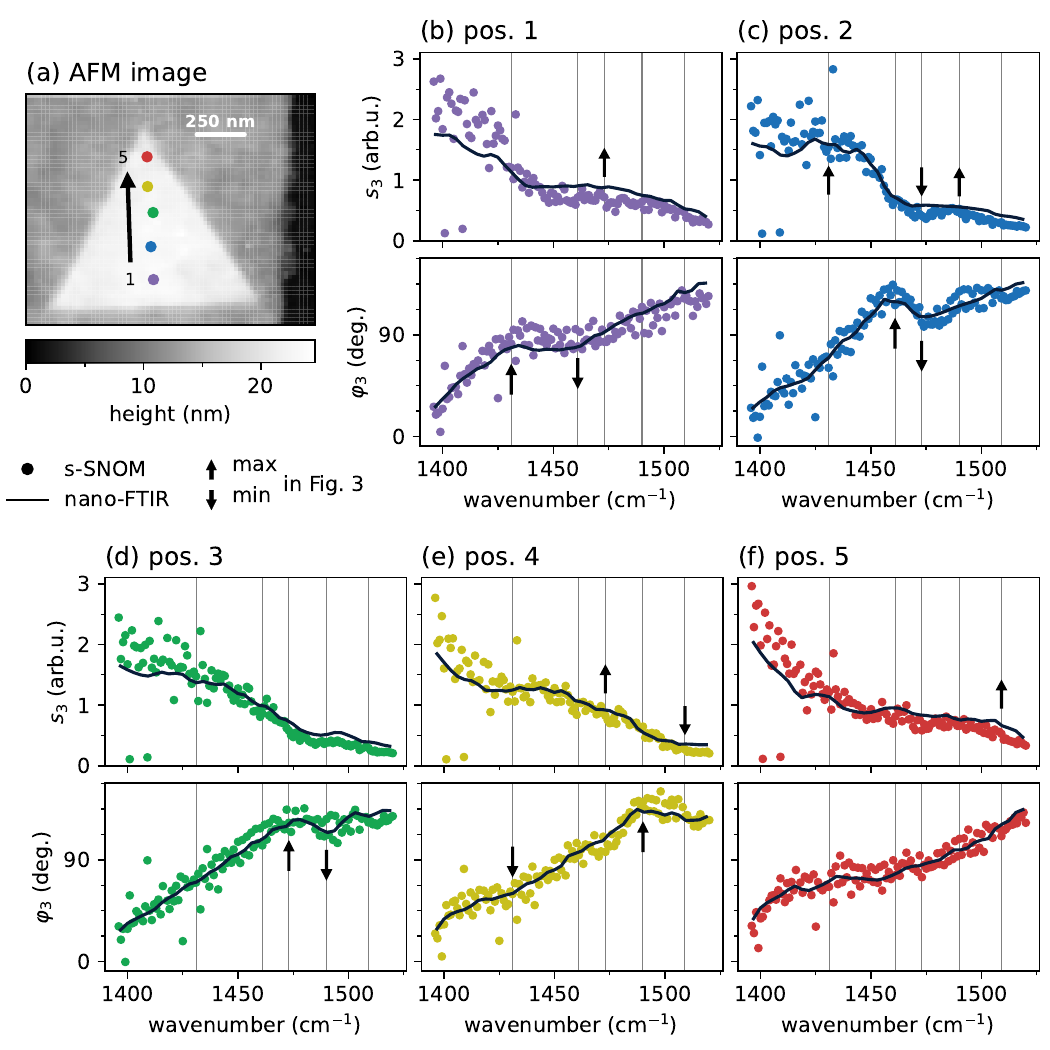}
\caption{s-SNOM and nano-FTIR spectroscopy of confined polariton modes. (a) AFM topography image of the $^{10}$BN triangular island (same as Figure~\ref{fig:2}(b)). Positions where spectra are taken are marked by colored dots. (b-c) nano-FTIR spectra (amplitude $s_3$ top and phase $\varphi_3$ bottom, solid black lines) taken at the positions marked in (a) together with the extracted s-SNOM values from the s-SNOM images in Figures~\ref{fig:3} and SI (colored dots). The gray vertical lines indicate the excitation wavenumbers corresponding to the s-SNOM maps shown in \ref{fig:3} and the arrows mark distinct maxima ($\uparrow$) and minima ($\downarrow$) at the respective position (colors) in the measured maps.}\label{fig:5}
\end{figure*}
%%%

Performing multi-frequency studies of s-SNOM maps renders a time-consuming task because one has to scan an entire sample area for each excitation wavenumber and afterwards align the images due to unavoidable sample drifts. Therefore, to retrieve entire spectra one typically performs nano-FTIR spectroscopy at distinct positions on the sample~\cite{huth2012nano,niehues2023identification}. However, in our situation the confined phonon polariton modes contain both spectral and spatial information. Therefore, we need to carry out nano-FTIR measurements at different locations in the sample to develop a comprehensive picture. To do so, we choose five different points along the symmetry line of the $^{10}$BN triangle as shown in the topography image in Figure~\ref{fig:5}(a)  (see SI for more measurement details). The measured nano-FTIR spectra are normalized to a Si sample and plotted for each position in Figures~\ref{fig:5}(b-f) as black curves (top panels nano-FTIR amplitude~$s_3$, bottom panels phase~$\varphi_3$). In addition, we plot the respective amplitudes and phases extracted from the 123 s-SNOM measurements evaluated at the same sample positions as dots (colors correspond to the ones in (a)). The normalization procedure as well as the spatial calibration algorithm are explained in the SI. We find that the overall agreement between nano-FTIR and s-SNOM spectra is remarkable.
In order to find a relation between features in these spectra and the selected s-SNOM images discussed before, we here marked the wavenumbers from Figure~\ref{fig:3} as vertical gray lines. As highlighted by the arrows, we find some correlations between spectral maxima ($\uparrow$) and minima ($\downarrow$) in the amplitude and phase spectra and spatial maxima or minima of $s_3$ of $\varphi_3$ in the s-SNOM maps in Figure~\ref{fig:3}. As an example we will discuss this correlation for position 2 (blue, Figure~\ref{fig:5}(c)) near the center of the triangle. The amplitude $s_3$ (top) has a spectral maximum at 1431~cm$^{-1}$ which is confirmed by the s-SNOM map in Figure~\ref{fig:3}(a) at the triangle center. Moving to 1461~cm$^{-1}$ in Figure~\ref{fig:5}(c) the phase spectrum $\varphi_3$ (bottom) has a clear maximum, which is in good agreement with the spatial maximum in Figure~\ref{fig:3}(b, bottom). For the next wavenumber 1473~cm$^{-1}$ $s_3$ and $\varphi_3$ have spectral minima in agreement with spatial minima at the triangle center in Figure~\ref{fig:3}(c). Finally, $s_3$ is maximal for 1490~cm$^{-1}$ and has a clear maximum at the center of the triangle in Figure~\ref{fig:3}(d, top). Overall we can summarize, that the mode confinement in the $^{10}$BN triangles is reflected by spectral features in point spectra. However, the significant spectral broadening of all confined modes, due to polariton decay and loose confinement, results in significant overlaps of all spectral resonances. Consequently, an identification of the triangle modes from the spectra alone is not possible and the characterization of the spatial field distribution via s-SNOM imaging is required.

\section{Conclusions}
We have thoroughly studied MOVPE-grown large-area BN layers regarding their phonon polariton properties and found that the epitaxial sample quality is high enough to detect polariton propagation for several wavelengths. While this improvement in MOVPE BN growth is an important progress by itself, our main achievement is the demonstration of phonon polariton confinement in MOVPE-grown $^{10}$BN micro-triangles. We have fully characterized the confined modes in these structures by s-SNOM and nano-FTIR measurements and modeling. Along the way, we have demonstrated that the two spectroscopy types can be directly compared not only qualitatively but also quantitatively. The demonstrated phonon polariton confinement renders an important progress in two different directions: (i) The engineering of van der Waals heterostructures provides many opportunities to tailor specific material properties~\cite{chen2023van}. Here, we have used an $^{10}$BN/$^{\text{nat}}$BN heterostructure to create phonon polariton resonators. (ii) The generation of micro-scaled high-quality polariton resonators provides opportunities to interface optically active lattice defects with phonons~\cite{wigger2019phonon,preuss2022resonant,niehues2025nanoscale}. This renders a promising perspective for novel quantum optical platforms~\cite{feng2026color}.\\
The overall excellent agreement between measured and simulated triangle modes allows for future design of desired phonon polariton confinement. Together with the demonstrated ability to grow high quality and large scale BN layers, this renders an important step towards the realization of scalable applications.
~\\[5mm]
\noindent\textbf{Acknowledgements} \par %delete if not applicable))
\noindent Ministerium für Innovation, Wissenschaft und Forschung des Landes Nordrhein-Westfalen: NRW-Rückkehr\-programm.
We thank the Münster Nanofabrication Facility (MNF) for providing access to the nano-FTIR setup.
This study was partially supported by the National Science Centre grants no. 2022/45/N/ST5/03396, 2022/47/B/ST5/03314, 2024/54/E/ST7/00273.

%\bibliography{refs}


\end{document}

\title{Phonon polariton confinement in isotopically pure MOVPE-grown BN triangles\\ {\it (Supporting Information)}}

\author{Maximilian~Scharpey\,\orcidlink{0009-0006-3642-6830}}
\affiliation{Institute of Physics, University of M\"unster, M\"unster, Germany}

\author{Oskar~Schr\"oer\,\orcidlink{0009-0006-5707-6569}}
\affiliation{Institute of Physics, University of M\"unster, M\"unster, Germany}

\author{Daniel Wigger\,\orcidlink{0000-0002-4190-8803}}
\affiliation{Department of Physics, University of M\"unster, Germany}

\author{Lena~Miler\,\orcidlink{0009-0002-3676-8129}}
\affiliation{Faculty of Physics, University of Warsaw, Warsaw, Poland}

\author{Jakub~Iwa{\'n}ski\,\orcidlink{0000-0003-2395-4010}}
\affiliation{Faculty of Physics, University of Warsaw, Warsaw, Poland}

\author{Andrzej~Wysmo\l{}ek\,\orcidlink{0000-0002-8302-2189}}
\affiliation{Faculty of Physics, University of Warsaw, Warsaw, Poland}

\author{Johannes~Binder\,\orcidlink{0000-0002-0461-7716}}
\affiliation{Faculty of Physics, University of Warsaw, Warsaw, Poland}

\author{Iris~Niehues\,\orcidlink{0000-0001-7438-2679}}
\email{iris.niehues@uni-muenster.de}
\affiliation{Institute of Physics, University of M\"unster, M\"unster, Germany}

%\email{email} 

\maketitle

\section{Sample growth} 
\begin{figure}
\centering
\includegraphics[width=0.5\textwidth]{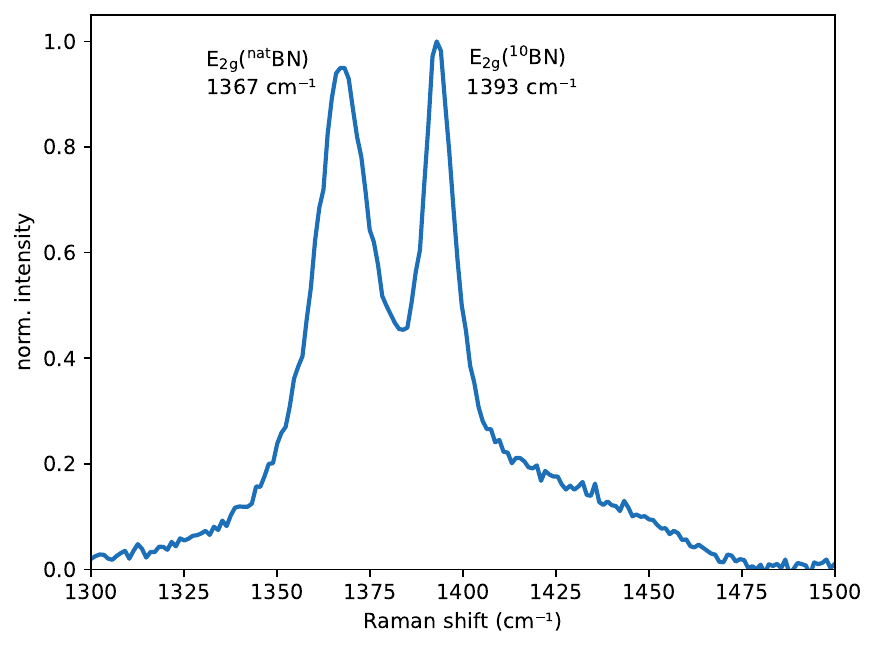}
\caption{Raman spectra measured on the $^{10}$BN/$^{\text{nat}}$BN triangular heterostructure showing the Raman peaks of both BN types.}\label{fig:S1}
\end{figure}

The samples were grown using an AIXTRON CCS 3×2” MOVPE system. The growth process followed the multistage protocol described in Ref.~\cite{dabrowska2020two}. Ammonia and triethylboron (TEB) were used as the nitrogen and boron precursors, respectively, with hydrogen serving as the carrier gas. The Continuous Flow Growth (CFG) stage was performed at an increasing temperature from 1250 to 1300 $^{\circ}$C, followed by Flow Modulation Epitaxy (FME) at 1400 $^{\circ}$C. The triangular islands were formed during the third step being CFG stage. For the $^{10}$BN triangular islands, the TEB precursor was switched to an isotopically purified source during the final growth step.
Figure \ref{fig:S1} shows the Raman spectra measured on the $^{10}$BN/natBN triangular heterostructure showing the Raman peaks of both BN types. The width of the E$_{2g}$ mode for $^{10}$BN is significantly smaller, confirming the absence of the $^{10}$B/$^{11}$B isotopic disorder present in natural BN.

\section{Experimental methods}
The s-SNOM/nano-FTIR measurements are executed using two different scattering-type optical near-field microscopy setups (\textit{Neascope} from \textit{Neaspec/Attocube})~\cite{keilmann2004near}. Both systems are based on an AFM operated in tapping mode (tip oscillation frequency $ \Omega\approx 260$~kHz). We use standard platinum-iridium tips (\textit{Arrow-NCPt} sourced from \textit{NanoWorld}) featuring a tip apex size of 30~nm as a near-field probe. Unwanted background signals are suppressed by demodulating the detector signal at the 3rd(4th) harmonic of the tip oscillating frequency, $3\Omega$($4\Omega$), yielding background-free amplitude and phase images/spectra, $s$ and $\varphi$, respectively, with nanoscale spatial resolution.\\
For the s-SNOM measurements (Figures~1 and 3, main text) a custom designed microscope equipped with a tunable monochromatic infrared QCL laser (1300~cm$^{-1}$ and 1700~cm$^{1}$, \textit{MIRcat-QT-Z-2200}) was used. The elastically backscattered light from the tip is recorded using an integrated pseudo-heterodyne interferometer~\cite{ocelic2006pseudoheterodyne}. Note, that this setup differs from a standard \textit{Neascope} due to a high-quality, silver protected off-axis parabolic mirror achieving a numerical aperture of $0.72$. In all s-SNOM measurements a tapping amplitude of approximately $80$~nm is chosen. For the s-SNOM measurements presented in Figure~1 in the main text, an excitation laser power of $1.5-2$~mW, a spatial sampling rate of 30~nm and an integration time of $13-16$~ms is used. The optical signal is analyzed in the 4th demodulation order.  In the large measurement series over 123 wavenumbers of the $^{10}$BN triangle (Figure~3, main text; SI video), a laser power of approximately 1.3~mW, a sampling rate of 20~nm, an integration time of 20~ms, and the third demodulation order is chosen.
The complex optical signal of all the resulting maps is normalized to the signal of the SiO$_{\mathrm{2}}$ substrate.\\
For the nano-FTIR spectroscopy measurements~\cite{amarie2009mid,huth2011infrared} shown in Figure~4 in the main text, we use a different \textit{Neascope} with the standard parabolic mirror with a numeric aperture of around 0.4. This setup is equipped with an integrated broadband laser light source (\textit{Toptica}), which emits light in the range of $1000~\mathrm{cm^{-1}}$ to $2000~\mathrm{cm^{-1}}$.
%\cite{amarie2009mid,huth2011infrared} 
The interferograms are recorded at a tapping amplitude of $70$~nm.  Each consists of $1024$ points recorded over a mirror distance of $800$~µm, where an integration time of $30$~ms is used for each point.
This results in a spectral resolution of $3.12~\mathrm{cm^{-1}}$.
To suppress the impact of noise for each measurement position on the triangle, we average over 30 single interferograms. The normalization of the spectra, is done using spectra recorded with the same measurement parameters but on a Si substrate. Before the measurement of each point-spectrum, an AFM topography map of the triangle is recorded to precisely set the desired measurement position and, in turn, compensate for unwanted tip misplacement resulting from sample drift.\\

\section{Phonon polariton s-SNOM measurements}
\begin{figure}
\centering
\includegraphics[width=0.5\textwidth, trim={0cm 2.5cm 5cm 0cm}, clip]{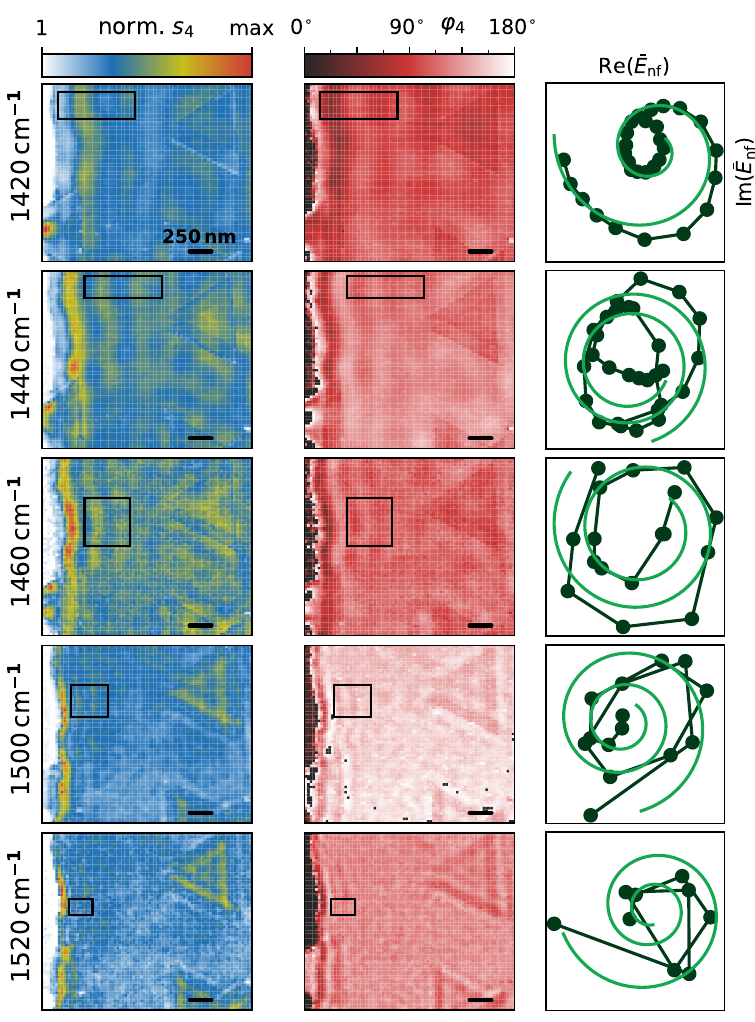}
\caption{s-SNOM maps and analysis to determine the $q$ vector for the dispersion relation in Figure~1 of the main text.}\label{fig:S2}
\end{figure}

Figure~\ref{fig:S2} shows s-SNOM images for five different excitation frequencies. To determine the corresponding polariton dispersion shown in Figure~1 in the main text, we select areas where the polariton waves resemble plane waves as best as possible. Note that this is more challenging than in high-quality crystals due to the slightly curved flake edge and the BN triangles which also act as scatterers for polariton waves. Consequently, the extracted data leading to spirals in the complex plane, also shown in the right column, are not particularly clean.
%%%
%%%

\section{Interference model for triangle modes}
\begin{figure}
\centering
\includegraphics[width=0.25\textwidth]{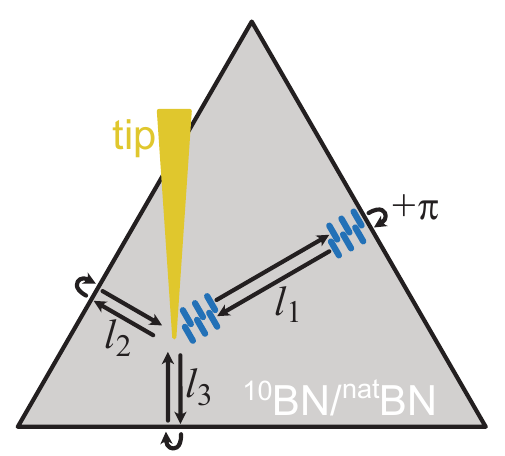}% wird noch neu gemacht 
\caption{Schematic picture of the interference model. Plane waves are launched from the tip and get reflected with a phase jump at all three edges. The resulting field at the tip position is detected.}\label{fig:S3}
\end{figure}

To simulate the phonon polariton waves confined in the triangle for a given excitation frequency $\omega$ we consider the scenario sketched in Figure~\ref{fig:S3}. At a given position of the AFM tip, the waves are launched in all in-plane directions. From these waves we also took into account that they travel to the three triangle edges and are reflected there (phase flip of $\pi$) and travel back to the tip. From the tip the resulting superposition of all fields, i.e., direct near fields and reflected ones, is scattered into the far field and detected. The total field then reads
\begin{align}
	E_{\rm nf} &= \sum_{j=1,2,3}  \big[1 + \exp\left(i2ql_j+i\pi\right)\exp\left(-2\gamma l_j\right)\big] \exp\left(-i\omega t\right)\,.
\end{align}
%%%
%%%

\section{Data analysis for the comparison of s-SNOM and nano-FTIR measurements}
To execute the comparison of the s-SNOM and nano-FTIR data (shown in Figure~4 in the main text), the data in the s-SNOM maps have to be analyzed at the exact position of the nano-FTIR spectra on the triangle.
To do so an algorithm was developed, which was able to find the position of the triangle's upper vertex in the topography maps. This algorithm was additionally used to compensate for spatial shifts between the s-SNOM maps in the SI video. In a first step a canny edge filter was used (Python package scikit-image, version 0.26.0, skimage.feature.canny) to detect any edges in the map. Using a Hough line transformation (Python package scikit-image, version $0.26.0$, skimage.transform.hough\_line), it was then possible to find the line parameters, which define the two triangle sides crossing in the upper vertex of the triangle. The position of the upper vertex in each AFM topography map was then calculated as the intersection point of the corresponding lines. Using this information, the relative vectors between each nano-FTIR measurement position and the upper triangle vertex was calculated. This set of vectors was then taken to the s-SNOM measurements, where they were added to the respective position of the upper vertex of each s-SNOM map. From here, the complex optical signal values of the pixel nearest to the resulting coordinates is extracted, which then were combined to form a spectrum complementary to the nano-FTIR results.\\
Due to different normalizations of the s-SNOM and nano-FTIR data, further normalization steps are necessary to allow for proper comparison. To do so, each amplitude spectrum (s-SNOM and nano-FTIR) was divided by its mean value, which then includes consideration of different bin numbers of the spectra. For the phase spectra, the difference between the mean values of corresponding s-SNOM and FTIR spectra was calculated. Finally, the resulting offsets were subtracted from the phase spectra of the FTIR measurement.

%\bibliography{refs}
